\documentclass[aps,prd,tightenlines,preprint,nofootinbib,a4paper]{revtex4}
\usepackage{epsfig}

\begin{document}

\title{Properties of the Neutrino Mixing Matrix}

\author{S. H. Chiu\footnote{schiu@mail.cgu.edu.tw}}
\affiliation{Physics Group, CGE, Chang Gung University, 
Kwei-Shan 333, Taiwan}

\author{T. K. Kuo\footnote{tkkuo@purdue.edu}}
\affiliation{Department of Physics, Purdue University, West Lafayette, IN 47907, USA}

\begin{abstract}

For neutrino mixing we propose to use the parameter set $X_{i}$ $(=|V_{ei}|^{2})$ and
$\Omega_{i}$ $(=\epsilon_{ijk}|V_{\mu j}|^{2}|V_{\tau k}|^{2})$,
with two constraints.  
These parameters are directly measurable since the neutrino oscillation
probabilities are quadratic functions of them.
Physically, the set $\Omega_{i}$ signifies a quantitative measure of $\mu-\tau$
asymmetry.  Available neutrino data indicate that all the $\Omega_{i}$'s are small
$(\lesssim O(10^{-1}))$, but with large uncertainties.  The behavior of $\Omega_{i}$
as functions of the induced neutrino mass in matter is found to be simple,
which should facilitate the analyses of long baseline experiments.

\end{abstract}


\maketitle

\pagenumbering{arabic}



\section{Introduction}

The recent results from the $\bar{\nu}$ disappearance experiments \cite{Daya,Reno} 
have made important contributions
toward pinning down the elements of the neutrino mixing (PMNS) matrix,
which we will denote as $V_{\nu}$, with elements $V_{\alpha i}$, $\alpha = (e,\mu,\tau)$, $i=(1,2,3)$.
Of the four physical parameters in $V_{\nu}$, three have now been measured, albeit with
different degrees of accuracy.  Furthermore, some information
on the fourth can already be gleaned from the known data set obtained by the various extant experiments,
as was done in global analyses thereof (see, \emph{e.g.}, \cite{Fogli,Forero}).  
It seems timely to study the available results in detail, with the intent to extract 
some general properties of $V_{\nu}$ which may be used to suggest directions 
for further investigation.

In this paper we will concentrate on two aspects of $V_{\nu}$.  First, it is interesting to assess
the impact of the known results on the possible symmetry properties of $V_{\nu}$.
Next, we address the behaviour of $V_{\nu}$ as a function of the parameter $A=2\sqrt{2}G_{F}n_{e}E$, 
the induced neutrino mass in matter.
This is especially relevant to the long baseline experiments (LBL)    
(for an incomplete list, see, \emph{e.g.},
Ref.\cite{CHA,LBL} and the references therein), which 
are generally regarded as the ``future'' of neutrino physics explorations.

The analyses of these issues are facilitated by a judicious choice of parameters for $V_{\nu}$.
To begin, in this study we will use the rephasing invariant parametrization
introduced earlier.  It consists of six parameters $(x_{i},y_{j})$, which satisfy two constraints. 
Prior to the recent measurement on $\bar{\nu}_{e} \rightarrow \bar{\nu}_{e}$, 
which results in a ``large'' $|V_{e3}|$,
it seems probable that $V_{\nu}$ is $\mu-\tau$ symmetric \cite{23-sym,LAM}, 
$|V_{\mu i}|=|V_{\tau i}|$.  
These conditions, when expressed in the variables $\Omega_{i}=x_{i}+y_{i}$, are just $\Omega_{i}=0$.
Now that $|V_{e3}|$ is non-vanishing, exact $\mu-\tau$ symmetry becomes less likely.
The question remains: `` Is there an approximate $\mu-\tau$ symmetry, and how good is it? ''.
It seems natural, then, to interpret $\Omega_{i}$ as the $\mu-\tau$ symmetry-breaking parameters.
As we shall see, they satisfy a constraint and thus there
are only two independent parameters in the set $\Omega_{i}$.
Also, taken together, the known neutrino data actually 
constrain all the $\Omega_{i}$'s so that none of which deviate much from zero.
The other physical parameters we choose are $X_{i}=x_{i}-y_{i}=|V_{ei}|^{2}$.  
As we shall see, the set $(X_{i},\Omega_{i})$ is
convenient for studying $V_{\nu}$, in vacuum as well as in matter.

For neutrino propagation in matter, it turns out that $X_{i}$, 
together with $\Delta_{ij}=m^{2}_{i}-m^{2}_{j}$,
satisfy a set of differential equations with respect to $A$, the induced mass.
Experimentally, the $X_{i}$'s and $\Delta_{ij}$'s are all very well measured in vacuum $(A=0)$.
This means that $X_{i}$ and $\Delta_{ij}$ are well determined, for all values of $A$.
Separately, $\Omega_{i}$ satisfy a set of differential equations containing $X_{i}$ and $\Delta_{ij}$.
They can be integrated to obtain $\Omega_{i}(A)$, with the known 
solution of $(X_{i},\Delta_{ij})$ as inputs.
Here, however, the initial values are poorly determined since there is only one direct measurement
(from atmospheric neutrinos) on $\Omega_{1}-\Omega_{2}$, with large errors.  
Nevertheless, it will be seen that
the $\Omega_{i}(A)$'s are largely constrained.

To go forward, our analysis suggests that the most urgent task would be another independent measurement
on $\Omega_{i}$.  It will clarify the nature of $V_{\nu}$ to a large extent.  At the same time, 
the close correlation between vacuum parameters and those in matter also means that LBL experiments
with variable $A$ can be extremely useful in the study of $V_{\nu}$.

\section{Rephasing invariant Parametrization}

As was shown before  \cite{Kuo:05,Chiu:09,CKL,CK10}, one can construct 
rephasing invariant combinations out of 
elements of a unitary, unimodular (det$V= +1$, so that $V^{*}$ = matrix of cofactors of $V$),
mixing matrix,
\begin{equation}\label{eq:g}
\Gamma_{ijk}=V_{1i}V_{2j}V_{3k}=R_{ijk}-iJ,
\end{equation}
where their common imaginary part can be identified with
the Jarlskog invariant $J$ \cite{Jar:85}.  
Their real parts are defined as
\begin{equation}
(R_{123},R_{231},R_{312};R_{132},R_{213},R_{321})
=(x_{1},x_{2},x_{3};y_{1},y_{2},y_{3}).
\end{equation}
These variables are bounded by $\pm 1$: 
$-1 \leq (x_{i},y_{j}) \leq +1$,
with $y_{j} \leq x_{i}$ for any ($i,j$). They satisfy two constraints
\begin{equation}\label{con1}
\mbox{det}V=(x_{1}+x_{2}+x_{3})-(y_{1}+y_{2}+y_{3})=1,
\end{equation}
\begin{equation}\label{con2}
(x_{1}x_{2}+x_{2}x_{3}+x_{3}x_{1})-(y_{1}y_{2}+y_{2}y_{3}+y_{3}y_{1})=0.
\end{equation}
In addition, it is found that 
\begin{equation}\label{eq:J}
J^{2}=x_{1}x_{2}x_{3}-y_{1}y_{2}y_{3}.
\end{equation}
Thus, the physical parameters contained in $V$ can be specified by the set
$(x,y)$ plus a sign, corresponding to $J=\pm\sqrt{J^{2}}$.

For applications to neutrino physics, it is traditional to label the matrix elements
as $V_{\alpha i}$, $\alpha=(e,\mu,\tau)$, $i=(1,2,3,)$.  The relations between 
$(x,y)$ and $|V_{\alpha i}|^{2}$ are given by
\begin{equation}\label{eq:w}
 W = \left[|V_{\alpha i}|^{2}\right]
   = \left(\begin{array}{ccc}
                    x_{1}-y_{1} & x_{2}-y_{2}   &  x_{3}-y_{3} \\
                     x_{3}-y_{2} & x_{1}-y_{3}  & x_{2}-y_{1} \\
                    x_{2}-y_{3}  &   x_{3}-y_{1}    & x_{1}-y_{2} \\
                    \end{array}\right).  
\end{equation}
One can readily obtain the parameters $(x,y)$ from $W$ by computing its cofactors,
which form the matrix $w$ with $w^{T}W=(\mbox{det}W)I$, and is given by
\begin{equation}\label{eq:co}
 w = \left(\begin{array}{ccc}
                    x_{1}+y_{1} & x_{2}+y_{2}   &  x_{3}+y_{3} \\
                     x_{3}+y_{2} & x_{1}+y_{3}  & x_{2}+y_{1} \\
                    x_{2}+y_{3}  &   x_{3}+y_{1}    & x_{1}+y_{2} \\
                    \end{array}\right).     
\end{equation}
Note that the elements of $w$ are bounded, $-1 \leq w_{\alpha i} \leq +1$, and
\begin{equation}
\sum_{i}w_{\alpha i}=\sum_{\alpha}w_{\alpha i}=\mbox{det} W,
\end{equation}
\begin{equation}
\mbox{det} W=\sum x^{2}_{i}-\sum y^{2}_{j}=\sum x_{i}+\sum y_{j},
\end{equation}
where the constraint equations Eq.~(\ref{con1}) and Eq.~(\ref{con2}) have been used.

The constraints Eqs.~(\ref{con1}) and (4) can be easily derived by using the identity
$\Gamma_{123}\Gamma_{231}\Gamma_{312}=\Gamma_{132}\Gamma_{213}\Gamma_{321}$.
One can obtain other useful relations when we consider product of the form
$\Gamma_{ijk}\Gamma_{lmn}$.  Thus, the well-known rephasing invariant expression
$\Pi^{\alpha \beta}_{ij}=V_{\alpha i}V_{\beta j}V_{\alpha j}^{*}V_{\beta i}^{*}$ consists of four
such terms.  For instance,
\begin{eqnarray}
\Pi^{e\mu}_{23}&=&(y_{1}-iJ)(y_{2}-iJ)-(x_{2}-iJ)(x_{3}-iJ) \nonumber \\
&-&(x_{1}-iJ)(x_{2}-iJ)+(x_{2}-iJ)(y_{3}-iJ).
\end{eqnarray}
The combination $\Pi^{\alpha \beta}_{ij}$ has the additional property that it is rephasing invariant even
if det$V=e^{i\theta} \neq 1$.  Another useful formula (with det$V=+1$) is
\begin{eqnarray}\label{Pi2}
\Pi^{\alpha \beta}_{ij} &=& |V_{\alpha i}|^{2}|V_{\beta j}|^{2}-
\sum_{\gamma k} \epsilon_{\alpha \beta \gamma}\epsilon_{ijk}V_{\alpha i}V_{\beta j}V_{\gamma k} \nonumber \\
&=& |V_{\alpha j}|^{2}|V_{\beta i}|^{2}+
\sum_{\gamma k} \epsilon_{\alpha \beta \gamma}\epsilon_{ijk}V^{*}_{\alpha j}V^{*}_{\beta i}V^{*}_{\gamma k},
\end{eqnarray}
where the second term in either expression is one of 
the $\Gamma$'s ($\Gamma^{*}$'s) defined in Eq.~(\ref{eq:g}).

We now turn to combinations of the form
$(y_{l}-iJ)(y_{m}-iJ)-(x_{i}-iJ)(x_{j}-iJ)$.
As an explicit example, consider
$(y_{1}-iJ)(y_{2}-iJ)-(x_{1}-iJ)(x_{2}-iJ) 
=(y_{1}y_{2}-x_{1}x_{2})+iJ(x_{1}+x_{2}-y_{1}-y_{2})$,
or
\begin{equation}\label{eq}
V_{e1}V_{e2}V_{e3}^{*}V_{\mu 3}V_{\tau 3}=(y_{1}y_{2}-x_{1}x_{2})+iJ(1-|V_{e3}|^{2}).
\end{equation}
In general, for $\alpha \neq \beta \neq \gamma$, $i\neq j\neq k$, 
\begin{equation}\label{general}
V_{\alpha j}V_{\alpha k}V_{\alpha i}^{*}V_{\beta i}V_{\gamma i}=
(y_{m}y_{n}-x_{b}x_{c})+iJ(1-|V_{\alpha i}|^{2}).
\end{equation}
Here, if $|V_{\alpha i}|^{2}=x_{a}-y_{l}$, then $b\neq c\neq a$, $m\neq n \neq l$.
Thus, if we take the matrix elements in the $\alpha$-th row and the $i$-th column,
complex conjugate the vertex $(V^{*}_{\alpha i})$, then the product is rephasing invariant
and has a well-defined imaginary part.  In fact, Eq.~(\ref{general}) provides another way to compute $J$.
For instance, in the standard parametrization \cite{data}, if we take $\alpha =e$ and $i=3$,
we quickly recover the usual expression for $J$.  The real part of Eq.~(\ref{general})
is also useful.  It enables us to compute other physical variables and, if $|V_{\alpha i}| \ll 1$,
set stringent bounds on them.
We will discuss these applications in sec. IV and V.


\section{Choice of variables}

Over the past couple of decades, a wealth of information has been gathered 
by neutrino oscillation experiments.
It would be useful to analyze the available data systematically so as to gain an overview 
of the neutrino mixing matrix.  To this end it is important to choose a set of parameters 
which can bring out clearly the salient features of $V_{\nu}$.  In this paper we propose to use certain
combinations of the variables $(x,y)$ which, as we shall see, can highlight the symmetry properties
of $V_{\nu}$.  In addition, they have simple behaviors when used in the study of 
neutrino propagation in matter.

Specifically, we choose the parameters,
\begin{equation}
X_{i}=x_{i}-y_{i}=|V_{ei}|^{2}=W_{ei},
\end{equation}
\begin{equation}\label{OI}
\Omega_{i}=x_{i}+y_{i}=\epsilon_{ijk}W_{\mu j}W_{\tau k}=w_{ei},
\end{equation}
where $i=(1,2,3)$.  Note that $-1 \leq \Omega_{i} \leq 1$, and
\begin{equation}
W_{\mu i}-W_{\tau i}=-\frac{1}{2}\epsilon_{ijk}(\Omega_{j}-\Omega_{k}),\
\end{equation}
\begin{equation}\label{4J}
4J^{2}=X_{1}X_{2}X_{3}+X_{1}\Omega_{2}\Omega_{3}+X_{2}\Omega_{1}\Omega_{3}+X_{3}\Omega_{1}\Omega_{2}.
\end{equation}
These variables are considered to be functions of $A=2\sqrt{2}G_{F}n_{e}E$, the induced neutrino mass,
which will be used when we discuss neutrino propagation in matter.
For the specific case of vacuum values, $A=0$, we will use the notation $X_{i}^{0}$ and $\Omega^{0}_{i}$.

The variables $X_{i}$ and $\Omega_{i}$ are not independent.  Obviously,
\begin{equation}\label{eq:A}
\sum X_{i}=1.
\end{equation}
The set $\Omega_{i}$ also satisfies a simple constraint.  From
\begin{equation}
W_{ei}w_{ei}=\mbox{det}W,
\end{equation}
and, using  Eqs.~(\ref{con1}) and (4),
\begin{equation}
\mbox{det} W=\sum x_{i}^{2}-\sum y_{j}^{2}=\sum x_{i}+\sum y_{j}=\sum \Omega_{i},
\end{equation}
we find
\begin{equation}\label{eq:B}
\sum \Omega_{i}(1-X_{i})=0
\end{equation}
The two constraints  Eqs.~(\ref{eq:A}) and (21) are equivalent to  Eqs.~(\ref{con1}) and (4),
but Eq. (21) is easier to implement since it is linear in $\Omega_{i}$.
Note also that if all the $\Omega_{i}$'s are equal, as happens when
$W_{\mu i}=W_{\tau i}$, then $\Omega_{i}=0$.

Thus, the rephasing invariant parametrization of $V_{\nu}$ consists of the set $(X_{i},\Omega_{i})$,
subject to two constraints,  Eqs.~(\ref{eq:A}) and (\ref{eq:B}).  Together with the mass differences,
$\Delta_{ij}=D_{i}-D_{j}$, $D_{i}=m_{i}^{2}$, they form a complete set of parameters 
for the neutrino oscillation phenomenology.

Before the recent measurements of $|V_{e3}|^{2}$, the possibility of a vanishing $|V_{e3}|^{2}$
and the equality $W_{\mu 3}=W_{\tau 3}$ led to the hypothesis of $\mu-\tau$ exchange symmetry for
neutrino mixing, $W_{\mu i}=W_{\tau i}$.  With the confirmed small, but non-vanishing, $|V_{e3}|^{2}$,
$\mu-\tau$ symmetry becomes less likely (although not excluded).  Nevertheless, it is 
interesting to analyze the property of $V_{\nu}$ under the $\mu-\tau$ exchange operation.
To this end, let us introduce a $\mu-\tau$ parity operator, $P_{\mu\tau}$, such that
\begin{equation}
P_{\mu\tau}:W_{\mu i} \leftrightarrow W_{\tau i}.
\end{equation}
From $X_{i}=1-(W_{\mu i}+W_{\tau i})$ and Eq.~(\ref{OI}), we find
\begin{equation}
P_{\mu \tau}: X_{i} \leftrightarrow X_{i},
\end{equation}
\begin{equation}
P_{\mu \tau}: \Omega_{i} \leftrightarrow -\Omega_{i},
\end{equation}
In addition,
since a $\mu-\tau$ exchange does not affect the eigenvalues of the neutrino mass matrix,
we also have
\begin{equation}
P_{\mu \tau}: \Delta_{ij} \leftrightarrow \Delta_{ij}.
\end{equation}
These results are independent of $A$, since the matter effect only contributes to the $e-e$ element
of the effective neutrino Hamiltonian.  Finally, from  Eq.~(\ref{4J}),
we see that $J^{2}$ is also invariant under $P_{\mu \tau}$, $J^{2} \leftrightarrow J^{2}$.

Thus, the neutrino parameters can be classified as 1) even under $P_{\mu \tau}$ : $X_{i}$,
$\Delta_{ij}$, $J^{2}$; 2) odd under $P_{\mu \tau}$ : $\Omega_{i}$.  The quantities
$\Omega_{i}$ serve as symmetry-breaking parameters -- they provide a measure of how good/bad
the $\mu-\tau$ exchange symmetry is.

We summarize our results in the matrix:
\begin{equation}\label{summary}
 W_{\nu} = \left(\begin{array}{ccc}
                    X_{1} & X_{2}   &  X_{3} \\
     \frac{1}{2}[(1-X_{1})+(\Omega_{3}-\Omega_{2})] & \frac{1}{2}[(1-X_{2})+(\Omega_{1}-\Omega_{3})]  & \frac{1}{2}[(1-X_{3})+(\Omega_{2}-\Omega_{1})] \\
     \frac{1}{2}[(1-X_{1})-(\Omega_{3}-\Omega_{2})] &  \frac{1}{2}[(1-X_{2})-(\Omega_{1}-\Omega_{3})]    & \frac{1}{2}[(1-X_{3})-(\Omega_{2}-\Omega_{1})] \\
                    \end{array}\right).     
\end{equation}

Eq.~(\ref{summary}) expresses the matrix $W_{ij}=|V_{ij}|^{2}$ interms of six parameters
($X_{i},\Omega_{i}$; with two constraints).  This may be contrasted with the set
$(\theta_{ij},\delta)$ of the standard parametrization.
While $V_{ij}$ is subject to rephasing, $|V_{ij}|^{2}$ quantifies the
physical mixing of states, and is directly measurable.  As we shall see
(Table 1), neutrino oscillations are simple functions of $(X_{i},\Omega_{i})$.
In terms of $(\theta_{ij},\delta)$, 
these functions become very complicated \cite{mnp}. 
In fact, one consequence is that there are multiple solutions of $(\theta_{ij},\delta)$,
corresponding to a given measurement.  
Thus, the set $(X_{i},\Omega_{i})$ offers a scheme which is
closely related to physical measurements.  And, it is hoped that 
the parameters can better quantify the nature of neutrino mixing.

Although in general the parameters $\Omega_{i}$ have the range, $-1 \leq \Omega_{i} \leq +1$,
given the current data, it will be seen (Sec. IV and V) that they are all small,
$|\Omega_{i}| \lesssim O(10^{-1})$, while vanishing values are not excluded.
Note also that the relations between $\Omega_{i}$ and the standard parameters
are given by \cite{CK11}:
\begin{eqnarray}
\Omega_{1} & = & c_{12}^{2}c_{13}^{2}(c_{23}^{2}-s_{23}^{2})-2K\cos\delta, \nonumber \\
\Omega_{2} &= &  -s_{12}^{2}c_{13}^{2}(c_{23}^{2}-s_{23}^{2})-2K\cos\delta, \nonumber \\
\Omega_{3} & = &  s_{13}^{2}(s_{12}^{2}-c_{12}^{2})(c_{23}^{2}-s_{23}^{2})
+2\frac{1+s_{13}^{2}}{1-s_{13}^{2}}K\cos\delta,
\end{eqnarray}
where $K=s_{12}c_{12}s_{13}c_{13}^{2}s_{23}c_{23}$.


\section{Neutrino mixing in vacuum}

Having settled on the parameter set $(X_{i},\Omega_{i})$, we turn now to the question of
their actual numerical values.  Since the experimental measurements have been given in terms
of the standard parametrization, we need to transcribe the results into the $(X_{i},\Omega_{i})$
variables.  In so doing, some informations are bound to be lost in translation.
Our numerical results are thus only approximate.  More precise ones can only be obtained by
analyzing directly the experiments in terms of the parameters $(X_{i},\Omega_{i})$.

For the actual numbers we will use the summaries from existing global analyses \cite{Fogli,Forero},
after making the proper conversion of variables.

First, the $X_{i}$'s are all well-determined.  There are slight differences between the two 
global analyses.  Also, the cases for normal and inverted mass spectra are not
significantly different.  We quote, approximately, 
$X^{0}_{2}=0.32 \pm 0.016$ \cite{Fogli};  $X^{0}_{2}=0.31 \pm 0.016$ \cite{Forero}, and   
$X^{0}_{3}=0.025 \pm 0.003$ \cite{Fogli}; $X^{0}_{3}=0.026 \pm 0.003$ \cite{Forero}. 
Of course, $X^{0}_{1}=1-X^{0}_{2}-X^{0}_{3}$.

Our knowledge on $\Omega^{0}_{i}$ is far less certain.  At the $1\sigma$ level, we have
$\Omega^{0}_{1}-\Omega^{0}_{2}=0.14$ to $0.24$ \cite{Fogli};  $= -0.14$ to $0.08$ \cite{Forero}. 
The discrepancy between these two results, as well as the large percentage errors in each,
is a reflection of the poor quality of them.

\begin{figure}[ttt]
\caption{Bounds (solid lines) from  Eq.~(\ref{omega3}) are plotted in the $(\Omega_{1}^{0},\Omega_{2}^{0})$
plane.  When combined with the $1\sigma$ bounds for $\Omega_{1}^{0}-\Omega_{2}^{0}$,
taken from \cite{Fogli} (dashed lines) and \cite{Forero} (dotted lines),
they indicate the allowed regions of $\Omega_{1}^{0}-\Omega_{2}^{0}$.
Note that the two regions do not overlap.} 
\centerline{\epsfig{file=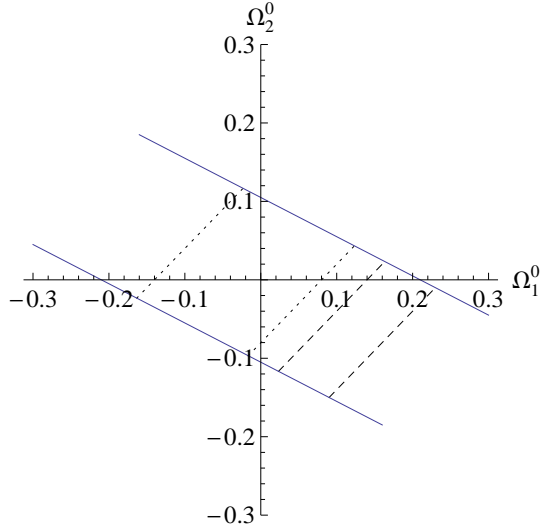,width= 7cm}}
\end{figure}

To complete the list we need one more piece of information on $\Omega^{0}_{i}$.
Despite the lack of another independent measurement, it turns out that the known
values of $X^{0}_{i}$ can already set a stringent bound on $\Omega^{0}_{3}$.  To see that we
return to Eq.~(\ref{eq}) in Sec II,
\begin{equation}\label{eq27}
V_{e1}V_{e2}V_{e3}^{*}V_{\mu 3}V_{\tau 3}=(y_{1}y_{2}-x_{1}x_{2})+iJ(1-|V_{e3}|^{2}). \nonumber
\end{equation}
This equation is especially useful if $|V_{e3}|^{2}=\epsilon^{2} \ll 1$.
In this case its LHS is significantly bounded since, in general, 
$|V_{e1}|^{2}|V_{e2}|^{2} \leq 1/4$ and $|V_{\mu 3}|^{2}|V_{\tau 3}|^{2} \leq 1/4$.
(\emph{E.g.}, 
$4|V_{e1}|^{2}|V_{e2}|^{2}=(|V_{e1}|^{2}+|V_{e2}|^{2})^{2}-(|V_{e1}|^{2}-|V_{e2}|^{2})^{2} \leq 1$).
Now,
\begin{equation}
2(y_{1}y_{2}-x_{1}x_{2})=-(X_{2}\Omega_{1}+X_{1}\Omega_{2}).
\end{equation}
Using vacuum values,
\begin{equation}
X^{0}_{2}\Omega^{0}_{1}+X^{0}_{1}\Omega^{0}_{2} \cong -\Omega^{0}_{3}.
\end{equation}
This follows from $\sum \Omega_{i}(1-X_{i})=0$, with the approximation
$X_{3} \cong 0$.  For $V_{\nu}$ in vacuum, actually $|V^{0}_{e1}|^{2}|V^{0}_{e2}|^{2}\cong 2/9$
and $|V^{0}_{\mu 3}|^{2}|V^{0}_{\tau 3}|^{2} \cong 1/4$.  Thus, with $X^{0}_{3}=\epsilon^{2}$,
\begin{equation}\label{JO}
(\frac{\Omega^{0}_{3}}{2})^{2}+(J^{0})^{2} \cong \epsilon^{2} /18,
\end{equation}
and
\begin{equation}\label{omega3}
|\Omega^{0}_{3}| \leq \sqrt{2}\epsilon /3 \cong 0.07.
\end{equation}
Also, in the same approximation,
\begin{equation}
\Omega^{0}_{1}+2\Omega^{0}_{2}+3\Omega^{0}_{3} \cong 0.
\end{equation}
The estimated values of $\Omega^{0}_{1}-\Omega^{0}_{2}$
can now be combined with the above bound 
($|\Omega^{0}_{1}+2\Omega^{0}_{2}| \leq 0.21$) 
in a plot in the ($\Omega^{0}_{1},\Omega^{0}_{2}$) plane, as shown in Fig. 1.
Here, the solid line correspond to the bound in Eq.~(\ref{omega3}).
We emphasize that this bound is robust, with possible deviations of no more than about $10\%$,
largely from the errors in $|V^{0}_{e3}|$.  On the other hand, the dashed \cite{Fogli} and
dotted \cite{Forero} lines have large errors, corresponding to the considerable
uncertainties in $\Omega^{0}_{1}-\Omega^{0}_{2}$. 
In this sense, Fig. 1 represents estimates of the probable values of 
$(\Omega_{1}^{0},\Omega_{2}^{0})$, but is not the traditional probability plot.
Nevertheless, the allowed regions for
$\Omega^{0}_{1}$ and $\Omega^{0}_{2}$ (and also $\Omega^{0}_{3}$) are essentially confined
to the neighborhood of the origin.  Despite the significant ambiguities, it seems that a
fair assessment is given by $|\Omega^{0}_{i}| \lesssim O(10^{-1})$.

In summary, given the incomplete results that are available at the present, one can
already deduce useful and quantitative information about all of the four physical parameters in
$(X^{0}_{i},\Omega^{0}_{i})$.  Clearly, the next step would be to have a precision measurement on
$\Omega^{0}_{i}$.  It is most useful to concentrate on $\Omega^{0}_{3}$, 
since this is equivalent to a measurement of $J^{0}$, according to Eq.~(\ref{JO}).

\begin{table*}[ttt]
 \centering
	\begin{center}
 \begin{tabular}{cccc}  
 
 \hline
\hline
   $Re(\Pi^{ee}_{21})$  &  & & $W_{e1}W_{e2}$    \\ 
   $Re(\Pi^{ee}_{31})$  & &  & $W_{e1}W_{e3}$   \\
    $Re(\Pi^{ee}_{32})$  & & &  $W_{e2}W_{e3}$  \\     
  $Re(\Pi^{ee}_{31}+\Pi^{ee}_{32})$  &  &  & $W_{e3}(1-W_{e3})$  \\       
    $Re(\Pi^{\mu\mu}_{21})$  &  & & $W_{\mu 1}W_{\mu 2}$ \\
   $Re(\Pi^{\mu \mu}_{31})$  &  & & $W_{\mu 1}W_{\mu 3}$ \\              
 $Re(\Pi^{\mu \mu}_{32})$  &  & & $W_{\mu 2}W_{\mu 3}$ \\
        $Re(\Pi^{\mu \mu}_{31}+\Pi^{\mu \mu}_{32})$  &  & & $W_{\mu 3}(1-W_{\mu 3})$  \\
   $Re(\Pi^{\mu e}_{21})$  & & &  $W_{e 1}W_{\mu 2}-x_{1}$ \\ 
  $Re(\Pi^{\mu e}_{31})$  &  & & $W_{e 1}W_{\mu 3}+y_{1}$ \\    
  $Re(\Pi^{\mu e}_{32})$  &  & & $W_{e 2}W_{\mu 3}-x_{2}$ \\
  $Re(\Pi^{\mu e}_{31}+\Pi^{\mu e}_{32})$  &  & & $-W_{e 3}W_{\mu 3}$ \\
      
      \hline
  \end{tabular}
    \caption{The amplitudes $Re(\Pi^{\alpha \beta}_{ij})$
are simple functions of $W_{\alpha i}$, or $X_{i}$ and $\Omega_{i}$.}
  \end{center}
 \end{table*}


Experimentally, the determination of $\Omega^{0}_{3}$ can be quite challenging.  First,
it is small $(\lesssim 0.07)$.  Second, its contribution to neutrino oscillation
probabilities is not easily disentangled.  We recall that the probability for
$\nu^{\alpha} \rightarrow \nu^{\beta}$ oscillation is given by
\begin{eqnarray}
P(\nu^{\alpha} \rightarrow \nu^{\beta})=\delta_{\alpha \beta}&-&4\sum_{i>j}Re(\Pi^{\alpha \beta}_{ij})\sin^{2}\Phi_{ij} \nonumber \\
&\pm &8(1-\delta_{\alpha \beta})J (\sin\Phi_{21}\sin\Phi_{31}\sin\Phi_{32}),
\end{eqnarray}
where $\Phi_{ij}=\Delta_{ij} L/4E$.  The amplitudes $Re(\Pi^{\alpha \beta}_{ij})$
are simple functions of $W_{\alpha i}$, or $X_{i}$ and $\Omega_{i}$, according to Eq.~(\ref{Pi2}),
and are listed in Table I.  In this table, it is sufficient to list $Re(\Pi^{\alpha \beta}_{ij})$
with $(\alpha, \beta)=(e,\mu)$.  This is because $\sum_{\alpha}Re(\Pi^{\alpha \beta}_{ij})=0$,
also $Re(\Pi^{\alpha \beta}_{ij})=Re(\Pi^{\beta \alpha}_{ij})$, since 
$\Pi^{\alpha \beta}_{ij}=(\Pi^{\beta \alpha}_{ij})^{*}$.
For vacuum values, $\Phi^{0}_{31} \simeq \Phi^{0}_{32}$, so we also list separately
the combinations $Re(\Pi^{\alpha \beta}_{31})+Re(\Pi^{\alpha \beta}_{32})$.

In searching for amplitudes that contain $\Omega_{3}$, it is clear that,
while $Re(\Pi^{\alpha \beta}_{31})$ or $Re(\Pi^{\alpha \beta}_{32})$ does
depend on $\Omega_{3}$, once we make the combination $Re(\Pi^{\alpha \beta}_{31}+\Pi^{\alpha \beta}_{32})$,
very few amplitudes have that property.  In fact, the six amplitudes of the form
$Re(\Pi^{\alpha \beta}_{21})$ and $Re(\Pi^{\alpha \beta}_{31}+\Pi^{\alpha \beta}_{32})$
fall into three groups. 1) $Re(\Pi^{ee}_{21})$ and $Re(\Pi^{ee}_{31}+\Pi^{ee}_{32})$:
They do not contain $\Omega_{i}$, and have been used to determine $X^{0}_{i}$ successfully.
2) $Re(\Pi^{\mu \mu}_{31}+\Pi^{\mu \mu}_{32})$ and $Re(\Pi^{\mu e}_{31}+\Pi^{\mu e}_{32})$.
Here, $W_{\mu 3}(1-W_{\mu 3}) \simeq \frac{1}{4}[1-(\Omega_{2}-\Omega_{1})^{2}]$,
and $-W_{e3}W_{\mu 3}\simeq -X_{3}[1+(\Omega_{2}-\Omega_{1})]/2$, for $X_{3}\ll 1$.
Thus, for vacuum values, they can only be used to determine $(\Omega^{0}_{2}-\Omega^{0}_{1})$.
Indeed, they, especially $W_{\mu 3}(1-W_{\mu 3})$, which gives the dominant contribution
to the atmospheric neutrino experiments, were used to infer that $(\Omega_{2}^{0}-\Omega_{1}^{0})$
is small.  At the same time, their structures also imply that very precise data are
needed in order to narrow down the errors of $(\Omega_{2}^{0}-\Omega_{1}^{0})$.
3) The amplitude $Re(\Pi^{\mu \mu}_{21})$ and $Re(\Pi^{\mu e}_{21})$ do depend on $\Omega_{3}$.
Substituting in the approximate vacuum values, $X^{0}_{1} \cong 2/3$, $X^{0}_{2} \cong 1/3$,
$X^{0}_{3}\approx 0$, and dropping quadratic terms in $\Omega_{i}$, we find
$Re(\Pi^{\mu \mu}_{21})^{0} \cong \frac{1}{18}-\frac{1}{6}(2\Omega^{0}_{2}+\Omega^{0}_{3})$,
$Re(\Pi^{\mu e}_{21})^{0} \cong -\frac{1}{9}+\frac{1}{6}(2\Omega^{0}_{2}+\Omega^{0}_{3})$.

Thus, to gain access to $\Omega^{0}_{3}$ (actually $\Omega^{0}_{3}+2\Omega^{0}_{2}$),
one has to isolate the amplitudes $Re(\Pi^{\mu \mu}_{21})^{0}$ and $Re(\Pi^{\mu e}_{21})^{0}$,
which is by no means easy.  We can only hope that the technical difficulties
involved can be overcome in the near future.


\section{Neutrino mixing in matter}

When neutrinos propagate in matter, their interactions induce a term in the effective Hamiltonian, given by
$(H)_{ee}=A=2\sqrt{2}G_{F}n_{e}E$ \cite{MSW}.  The mass eigenvalues and the mixing matrix are now functions of
$A$.  It was shown \cite{CK11} that they satisfy a set of differential 
equations listed in Table II, together with
\begin{equation}\label{V1}
\frac{dD_{i}}{dA}=X_{i}.
\end{equation}

\begin{table*}[ttt]
 \centering
	\begin{center}
 \begin{tabular}{cccc}  
 
 \hline

    & $1/\Delta_{12}$  & $1/\Delta_{23}$     &  $1/\Delta_{31}$       \\ \hline
   $\frac{1}{2}\frac{d}{dA}\ln X_{1}$ &  $X_{2}$  &    &  $-X_{3}$ \\
   $\frac{1}{2}\frac{d}{dA}\ln X_{2}$ & $-X_{1}$ & $X_{3}$   &   \\     
  $\frac{1}{2}\frac{d}{dA}\ln X_{3}$ &  & $-X_{2}$  & $X_{1}$  \\       
   $\frac{d\Omega_{1}}{dA}$ & $\Omega_{1}X_{2}-\Omega_{2}X_{1}$  & 
     $-\Omega_{1}X_{2}-\Omega_{2}X_{1}+\Omega_{1}X_{3}+\Omega_{3}X_{1}$  &$-\Omega_{1}X_{3}+\Omega_{3}X_{1}$ \\              
  $\frac{d\Omega_{2}}{dA}$ &  $\Omega_{1}X_{2}-\Omega_{2}X_{1}$  & $\Omega_{2}X_{3}-\Omega_{3}X_{2}$
          & $\Omega_{1}X_{2}+\Omega_{2}X_{1}-\Omega_{2}X_{3}-\Omega_{3}X_{2}$ \\
  $\frac{d\Omega_{3}}{dA}$ & $-\Omega_{1}X_{3}-\Omega_{3}X_{1}+\Omega_{2}X_{3}+\Omega_{3}X_{2}$  
   & $\Omega_{2}X_{3}-\Omega_{3}X_{2}$  &  $-\Omega_{1}X_{3}+\Omega_{3}X_{1}$  \\         
      \hline
  \end{tabular}
    \caption{The differential equations for $X_{i}$ and $\Omega_{i}$ in matter, expressed as the sums of terms
    proportional to $1/\Delta_{ij}$.}
  \end{center}
 \end{table*}

These equations are derived from
\begin{equation}
(V+dV)^{\dag}(H+dH)(V+dV)=D+dD,
\end{equation}
which is just the flavor-basis version of the familiar perturbation theory in quantum mechanics,
\begin{equation}
(1+dV)^{\dag}(H+dH)(1+dV)=D+dD,
\end{equation}
written in the mass eigenstate basis.  Thus, the quantum mechanical result $dD_{i}=\langle i|dH|i\rangle$
becomes Eq.~(\ref{V1}), $dD_{i}=|V_{ei}|^{2} dA$, when $dH$ has only an $(ee)$ element,
$\langle e|dH|e\rangle =dA$.
Similarly, the equations for $X_{i}$ and $\Omega_{i}$ are just rephasing invariant combinations
constructed out of the well-known formula $dV_{ij}=\langle i|dH|j\rangle /(D_{j}-D_{i})$,
after its conversion into the flavor basis. 

The above arguments should help to illuminate the nature of these
differential equations and to put on firm grounds the use of them
in solving the mixing problem.
It is noteworthy that the equations for $D_{i}$ and $X_{i}$ are independent of $\Omega_{i}$.
These equations are even under $P_{\mu \tau}$ so that, barring the possible appearance of higher order terms,
the $\Omega$'s are absent.  Similarly, the differential equations for $\Omega_{i}$ are odd under $P_{\mu \tau}$,
so that they are all linear in $\Omega_{i}$.
Previously \cite{CK11}, we studied the $A$-dependence of $X_{i}$ and $D_{i}$, assuming $\Omega_{i}=0$.
We now see that even with $\Omega_{i} \neq 0$, the conclusions remain valid.  We only need to add
$\Omega_{i}$ to the list and find out how they behave.

Let us first summarize the results for $X_{i}$ and $\Omega_{i}$.
Because the mass differences $\Delta^{0}_{21}$ and $\Delta^{0}_{32}$ are widely separated, it is
known (see \emph{e.g.}, Ref. \cite{KP}) that the three-flavor problem can be well approximated 
by two, two-flavor, level crossing problems, occurring at $A=A_{l}$, when
$\Delta_{21}=min$, and at $A=A_{h}$, for $\Delta_{32}=min$.
In our present formulation this is just the pole dominance approximation. 
Near $A \sim A_{l}$, the differential equations for $(D_{1},D_{2},X_{1},X_{2})$
are reduced to the pole term $\propto 1/D_{21}$, plus Eq.~(\ref{V1}).
Similarly, for $A \sim A_{h}$, variations of $(D_{2},D_{3},X_{2},X_{3})$ are governed by Eq.~(\ref{V1})
and the pole term $\propto 1/D_{32}$.  Consequently, the function $D_{i}(A)$ takes turn to rise linearly,
while $X_{i}$ behaves as step-functions.  We list these results by dividing $A$ into three regions,

I) $A \lesssim A_{l}$: $D_{1} \propto X^{0}_{1}A$, $D_{2} \propto X^{0}_{2}A$, $D_{3} \cong \mbox{constant}$,
$X_{i} \cong X^{0}_{i}$; \\

II) intermediate $A_{i}$ region, $A_{l} \lesssim A \lesssim A_{h}$: $(D_{1},D_{3}) \cong \mbox{constant}$,
$D_{2} \propto A$, $X_{1} \rightarrow 0$, $X_{2} \rightarrow 1$, $X_{3} \cong X^{0}_{3}$; \\

III) dense medium $A_{d}$, $A \gtrsim A_{h}$: $(D_{1},D_{2}) \cong \mbox{constant}$,
$D_{3} \propto A$, $(X_{1},X_{2})\rightarrow 0$, $X_{3} \rightarrow 1$. \\
Detailed graphs for these variables will be given in Sec. VI.

\begin{figure}[ttt]
\caption{The parameters $X_{i}$ (left column) and $\bar{X}_{i}$ (right column) as functions of $A/\Delta^{0}_{21}$
for both the normal (solid) and the inverted (dashed) mass spectra.
The initial values in vacuum are well-measured, 
with $X^{0}_{1}=\bar{X}^{0}_{1}\simeq (2/3)(1-X^{0}_{3})$,  
$X^{0}_{2}=\bar{X}^{0}_{2}\simeq (1/3)(1-X^{0}_{3})$, 
where $X^{0}_{3}=\bar{X}^{0}_{3}=|V^{0}_{e3}|^{2}$.} 
\centerline{\epsfig{file=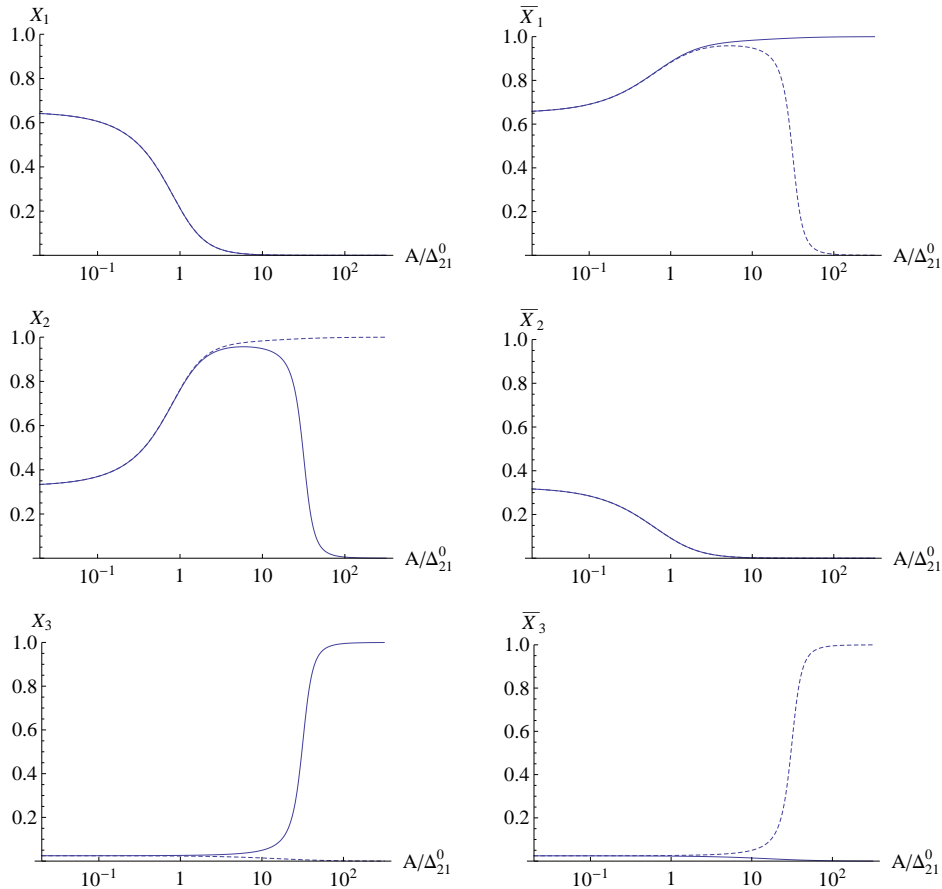,width=12.5 cm}}
\end{figure}

\begin{figure}[ttt]
\caption{The qualitative plots for $D_{1}$ (dashed), $D_{2}$ (solid),
and $D_{3}$ (dot-dashed) under normal (left) and inverted (right) hierarchies.
Note that in each plot, the curves in the positive region of $A$ represent $D_{i}$ for
the $\nu$-sector, while that in the negative region of $A$ represent $D_{i}$ for the $\bar{\nu}$-sector.
} 
\centerline{\epsfig{file=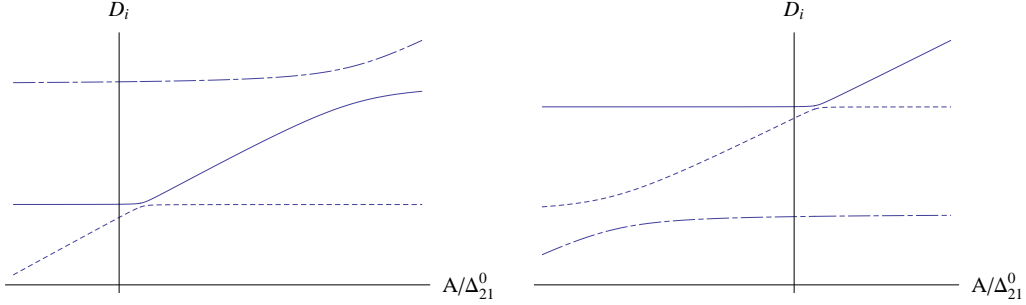,width=13.5 cm}}
\end{figure} 

Having obtained the functions $D_{i}(A)$ and $X_{i}(A)$, we can now use them to evaluate $\Omega_{i}(A)$.
While one can integrate the differential equations for $\Omega_{i}$ (Table II) numerically,
which will be presented in Sec. VI, much of the behavior of $\Omega_{i}(A)$ can already be inferred
from the constraints (Eq.~(\ref{eq:B})) and the nature of the differential equations they satisfy.
As with $X_{i}(A)$ and $D_{i}(A)$, we will now invoke the pole dominance approximation, 
so that $\Omega_{i}(A)$ undergo rapid variations only near the resonances $A \sim A_{l}$ and $A \sim A_{h}$.
Outside of these, $\Omega_{i}(A)$ are essentially flat.  Their values can be summarized as follows. 

I) In the intermediate $A$ region, $A_{l} \lesssim A \lesssim A_{h}$: $\Omega_{1}\rightarrow 0$,
$\Omega_{2} \rightarrow \Omega^{0}_{2}-\Omega^{0}_{1}$, $\Omega_{3} \rightarrow 0$.
Here, $X_{1} \rightarrow 0$, $X_{2} \rightarrow 1-X^{0}_{3} \cong 1$, $X_{3} \rightarrow X^{0}_{3}$.
Also, a more precise $X_{1}(A)$ can be obtained by using 
$X^{0}_{1}X^{0}_{2}(\Delta^{0}_{21})^{2}=X_{1}X_{2}(\Delta_{21})^{2}$ (Eq. (4.12) of Ref. \cite{CK10}).
With $X_{2} \rightarrow 1$, $(\Delta_{21})^{2} \cong A^{2}$, and $X^{0}_{1}X^{0}_{2} \cong 2/9$,
we have $X_{1}(A) \cong (2/9)(\Delta^{0}_{21}/A)^{2}$, for $A_{l} \lesssim A \lesssim A_{h}$.
$X_{1}(A)$ is thus already very small for $A \gtrsim 5\Delta^{0}_{21}$.
Now we turn to the identity Eq.~(\ref{general}), for $\alpha=e$, $l=1$,
\begin{equation}
V_{\mu 1}V_{\tau 1}V_{e1}^{*}V_{e2}V_{e3}=(-x_{2}x_{3}+y_{2}y_{3})+iJ(1-|V_{e1}|^{2}).
\end{equation}
With $|V_{e1}|=\epsilon' \ll 1$, $|V_{\mu 1}V_{\tau 1}| \leq 1/2$, $|V_{e3}|=\epsilon \ll1$,
and $\Omega_{1} \cong -(\Omega_{2}X_{3}+\Omega_{3}X_{2})=-2(x_{2}x_{3}-y_{2}y_{3})$,
we find the bound
\begin{equation}
|\Omega_{1}| \leq \epsilon \epsilon',
\end{equation}
for $A_{l} \lesssim A \lesssim A_{h}$.  At the same time, the constraint $\sum \Omega_{i}(1-X_{i})=0$,
after putting in the values of $X_{i}$, becomes
\begin{equation}
\Omega_{1}+\Omega_{3} \cong 0.
\end{equation}
Finally, referring to Table II, we have
\begin{equation}
\frac{d}{dA}(\Omega_{2}-\Omega_{1})=0,
\end{equation}
at the $A_{l}$ pole.  Thus,
\begin{equation}
\Omega^{0}_{2}-\Omega^{0}_{1}=\Omega_{2}-\Omega_{1} \rightarrow \Omega_{2}(A)
\end{equation}
for $A_{l} \lesssim A \lesssim A_{h}$. 

II) For dense medium, $A \gtrsim A_{h}$: $\Omega_{1} \rightarrow 0$, $\Omega_{2} \rightarrow 0$, 
$\Omega_{3} \rightarrow -\Omega^{0}_{2}+\Omega^{0}_{1}$.  In this region, $X_{1} \rightarrow 0$,
$X_{2} \rightarrow 0$, $X_{3} \rightarrow 1$.  Following steps as above, 
we see first that $\Omega_{2} \rightarrow 0$.  The constraint equation then gives
$\Omega_{1}+\Omega_{2}=0$, and, using the property of the pole at $A_{h}$,
$d(\Omega_{3}-\Omega_{2})/dA=0$, we find the results listed.  Note that the initial conditions
for the differential equation are $\Omega_{3}(A_{i})=0$, $\Omega_{2}(A_{i})=\Omega^{0}_{2}-\Omega^{0}_{1}$.

Our results show that $\Omega_{i}(A)$ behave rather simply as functions of $A$.
The transition regions near $A_{l}$ and $A_{h}$, however, are not covered.
For these we need to solve the differential equations numerically, which will be given in Sec. VI.
After these transition regions, the resonances act to ``quench'' many of the parameters.
Thus, as $A$ increases, much of the information in the original set $(X^{0}_{i},\Omega^{0}_{i})$
will disappear, and $W_{\nu}$ contains fewer and fewer free parameters.
Physically, this is a consequence of the decoupling of the $|e\rangle$ state, 
whose mass rises roughly as a linear function of $A$.  Starting as a mixture of
$|1\rangle$ and $|2\rangle$, it evolves into an almost pure $|2\rangle$ state,
and eventually becomes just a $|3\rangle$ state.
For $A \gg m^{2}_{3}$, the system is made up of $|e\rangle$ ($|3\rangle$) plus a completely decoupled 
two-flavor system, consisting of $|\mu\rangle$ and $|\tau\rangle$, or $|1\rangle$ and $|2\rangle$.
The vestige of the original $\mu-\tau$ mixing, $(\Omega^{0}_{2}-\Omega^{0}_{1})$,
however, remains undisturbed by the change in the $e-$channel, coming from $\Delta H_{ee}$.

So far we have tacitly assumed the ``normal'' ordering of the neutrino masses.
For the case of ``inverted'' ordering, the resonance at $A_{h}$ is absent, but the 
rest of our analyses remain valid.

\begin{figure}[ttt]
\caption{The numerical solutions of $\Omega_{i}$ for the $\nu$-sector 
under both the normal hierarchy (solid) and the inverted hierarchy (dashed).
Two different sets of initial values are adopted: $(\Omega^{0}_{1},\Omega^{0}_{2},\Omega^{0}_{3})=(0.18,-0.015,-0.05)$ \cite{Fogli} 
(left column) and $(\Omega^{0}_{1},\Omega^{0}_{2},\Omega^{0}_{3})=(0.06,0.045,0.05)$ \cite{Forero} 
(right column).} 
\centerline{\epsfig{file=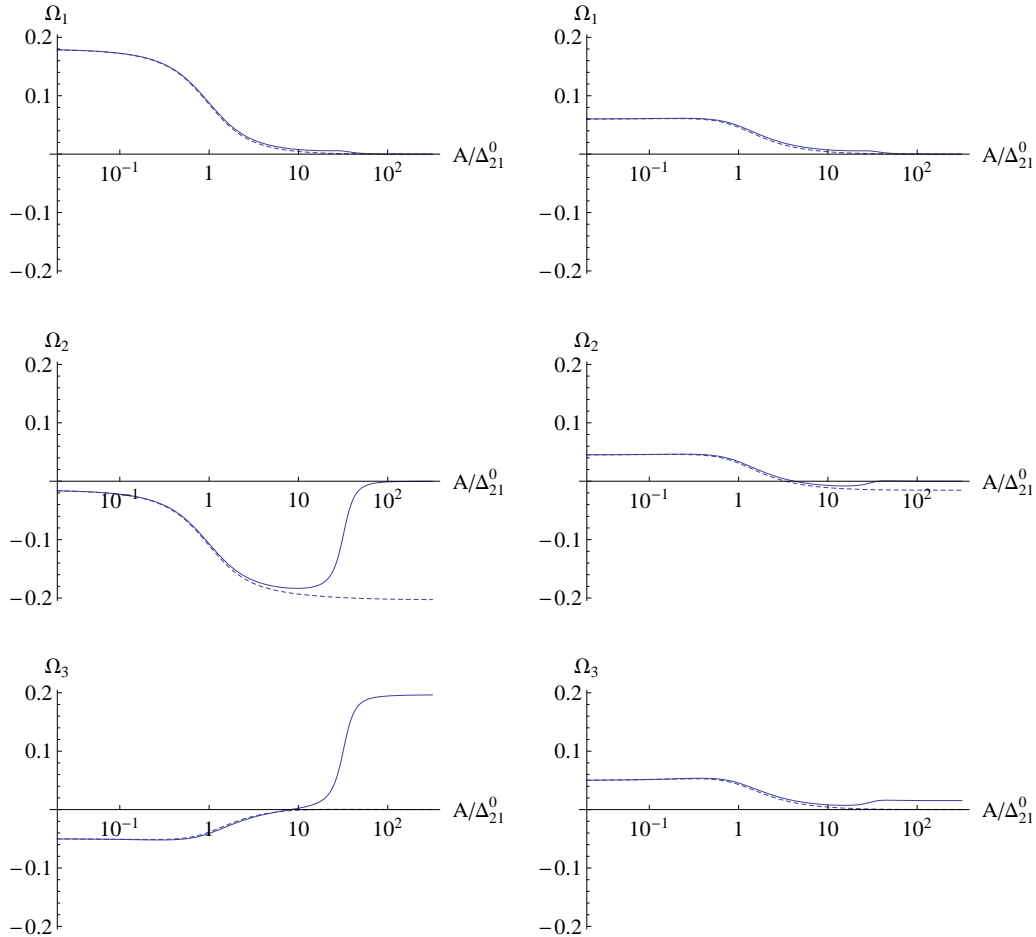,width=13.5 cm}}
\end{figure}

\begin{figure}[ttt]
\caption{The numerical solutions of $\bar{\Omega}_{i}$ for the $\bar{\nu}$-sector
under both the normal hierarchy (solid) and the inverted hierarchy (dashed).
We adopt two different sets of initial values: $(\Omega^{0}_{1},\Omega^{0}_{2},\Omega^{0}_{3})=(0.18,-0.015,-0.05)$ \cite{Fogli} 
(left column) and $(\Omega^{0}_{1},\Omega^{0}_{2},\Omega^{0}_{3})=(0.06,0.045,0.05)$ \cite{Forero} 
(right column).} 
\centerline{\epsfig{file=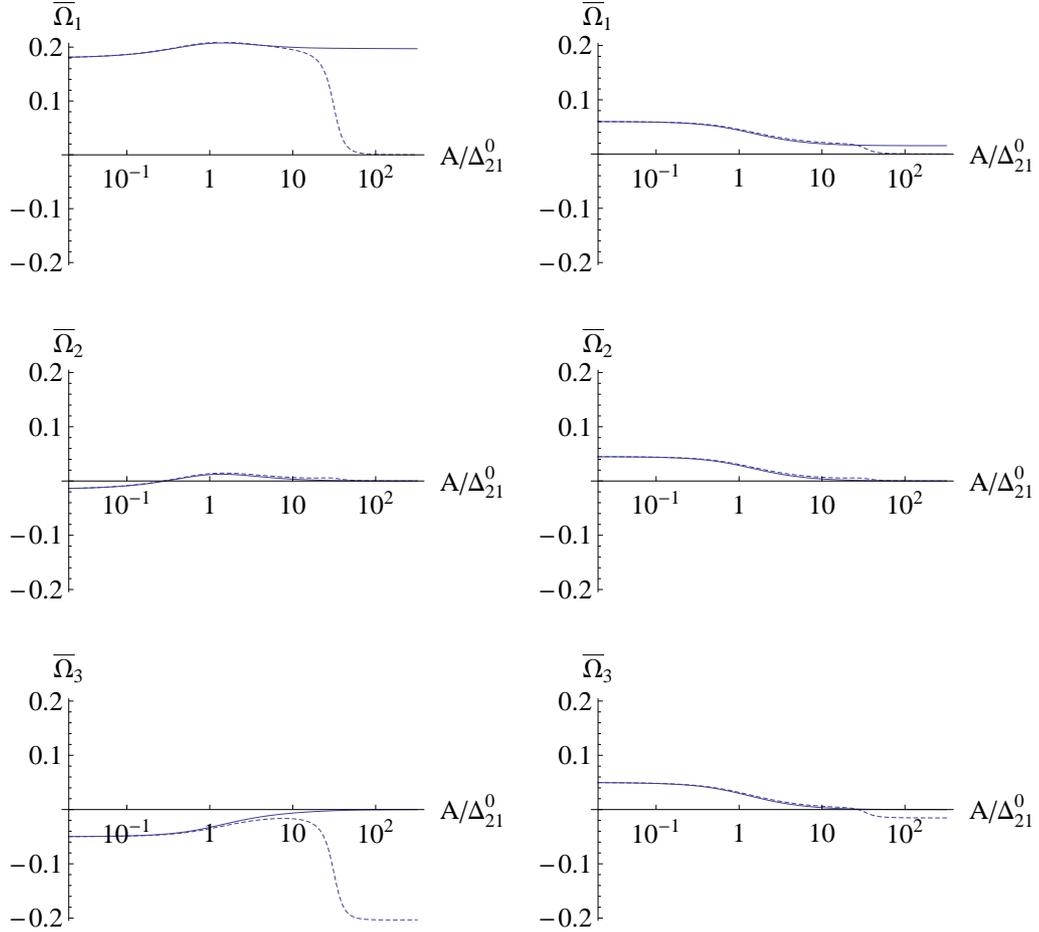,width=13.5 cm}}
\end{figure}


\section{Numerical solutions}


We summarize our analyses of the parameters here with their numerical solutions.
As indicated earlier, the quantities $X^{0}_{i}$ and $\Delta^{0}_{ij}=D_{i}-D_{j}$ are
all well measured at $A=0$, with errors $\sim 10\%$.
We show both $X_{i}$ (for $\nu$) and $\bar{X}_{i}$ (for $\bar{\nu}$) 
as functions of $A/\Delta^{0}_{21}$ in Fig. 2
with the initial values:
$X^{0}_{1}=\bar{X}^{0}_{1}\simeq (2/3)(1-X^{0}_{3})$ and  $X^{0}_{2}=\bar{X}^{0}_{2}\simeq (1/3)(1-X^{0}_{3})$, 
where $X^{0}_{3}=\bar{X}^{0}_{3}=|V^{0}_{e3}|^{2}$
is taken from both Refs. \cite{Fogli} and \cite{Forero}.
It is seen that the paths for
$\nu$ and $\bar{\nu}$ begin to evolve apart when $A \gtrsim A_{l}$.
In addition, $X_{1}$ and $\bar{X}_{2}$ are insensitive to the mass hierarchy, while each of
$X_{2}$, $X_{3}$, $\bar{X}_{1}$, and $\bar{X}_{3}$ evolves diversely under different mass hierarchies
when $A>A_{l}$, where the higher resonance begins to affect.  
Note that with the exchange: normal $\leftrightarrow$ inverted, 
the ``trends''  of the following curves evolve similarly:
$X_{1} \sim \bar{X}_{2}$, 
$X_{2} \sim \bar{X}_{1}$, and $X_{3} \sim \bar{X}_{3}$.
The evolution details of $X_{i}$ and $\bar{X}_{i}$
can be readily inferred from how $D_{i}$ varies, as indicated by Eq.~(\ref{V1}).
As an illustration, we provide the well-known qualitative plots of $D_{i}$ in Fig. 3.

On the other hand,
$\Omega^{0}_{i}$ are only roughly known from global fits, with typical errors of
$30\sim 100\%$.
The evolution equations for $\Omega_{i}$ in Table II can be expressed in a compact form,
\begin{eqnarray}\label{eq:Omega}
\frac{d\Omega_{i}}{dA}&=&\sum_{j>k}\frac{1}{\Delta_{jk}}[\delta_{ij}(\Omega_{i}X_{k}-\Omega_{k}X_{i})
-\delta_{ik}(\Omega_{i}X_{j}-\Omega_{j}X_{i})] \nonumber \\
&+&\sum_{j>k,i \neq j\neq k}\frac{1}{\Delta_{jk}}[-(\Omega_{i}X_{j}+\Omega_{j}X_{i})+(\Omega_{i}X_{k}+\Omega_{k}X_{i})].
\end{eqnarray}
The numerical solutions of $\Omega_{i}$ for both mass spectra are shown in Fig. 4.
It is seen that the evolution of $\Omega_{i}$ is sensitive to the choice of initial values,
which are not quite settled experimentally. 
The general features, however, can be fairly understood from the constraint, Eq.~(\ref{eq:B}),
and the nature of the evolution equations, as discussed in Sec. V. 
Note that the curves for the inverted mass ordering follow more closely the trend we discussed in Sec. V
since they are not distorted by the higher resonance.

For the $\bar{\nu}$-sector, the evolution of $\bar{\Omega}_{i}$ in matter is shown in Fig. 5.
The analyses in Sec. IV remain valid to the understanding of their general features.
Note that $\Omega_{i}$ ($\bar{\Omega}_{i}$)  
share similar qualitative properties as $X_{i}$ ($\bar{X}_{i}$),
$e.g.$, $\Omega_{1}$ and $\bar{\Omega}_{2}$ are insensitive to the mass hierarchy even 
when the neutrinos propagate in matter, while the matter effect 
breaks the hierarchy degeneracy 
for each of the parameters, $\Omega_{2}$, $\Omega_{3}$, $\bar{\Omega}_{1}$,
and $\bar{\Omega}_{3}$, when $A>A_{l}$.


As one scans through the plots, it is noteworthy that all of the parameters 
undergo rapid changes
near the resonance positions, but are otherwise more or less flat. 
This lends support to the use of the pole approximation in Sec. V.  The values
before and after the transition regions also agree with the estimates
given in Sec. V.  For instance, consider Fig. 4.  Here, $\Omega_{1}$
drops from $\Omega^{0}_{1} =0.18$ to almost zero, when $A/\Delta^{0}_{21}$ varies from about 0.5 to 3.  
At the same time, $\Omega_{3}$ changes from -0.05 to $\sim 0$, while $\Omega_{2}$
goes from -0.01 to $\sim -0.18$, not far from $\Omega^{0}_{2}-\Omega^{0}_{1} \approx -0.19$. 
From $A/\Delta^{0}_{21} \sim 3$ to $\sim 25$, $\Omega_{i}$ stays nearly constant.
The other plots can be similarly analyzed.  To summarize, unless very high
accuracy is demanded, the approximation presented in Sec. V should be
sufficient for most purposes.


\section{conclusion}

The physics of neutrino oscillation is governed by a mixing matrix $V_{\nu}$ which
is unitary, unimodular, and rephasing invariant.  As such $V_{\nu}$ (or $W_{\nu}$)
satisfies a number of self-consistency conditions (such as Eq.~(\ref{general})) 
which can help
to characterize the matrix.  To exploit these properties,
we propose to use the parameters $X_{i}$ ($=W_{ei}$) and 
$\Omega_{i}$ 
(with $W_{\mu i}-W_{\tau i}=\frac{1}{2}\epsilon_{ijk}(\Omega_{k}-\Omega_{j})$), 
which satisfy two constraints, 
Eq.~(\ref{eq:A}) and Eq.~(\ref{eq:B}).
Physically, the set $\Omega_{i}$ offers a measure of the $\mu-\tau$ asymmetry.
These parameters are directly measurable, since the neutrino oscillation
probabilities $P(\nu^{\alpha}\rightarrow \nu^{\beta})$ are simple functions of them.
This parametrization is summarized in Eq.~(\ref{summary}).

Experimentally, the vacuum values of $X_{i}$ ($X^{0}_{i}$) are well-determined.
But for $\Omega_{i}$ there is only one measurement on $(\Omega^{0}_{2}-\Omega^{0}_{1})$,
which turns out to be small ($O(10^{-1})$), with large uncertainties.
However, using Eq.~(\ref{eq27}), one can establish the bound $\Omega^{0}_{3} \lesssim 0.07$,
whose validity depends on the approximation $X^{0}_{3} \ll 1$.
There is also a sum rule relating $\Omega^{0}_{3}$ to $J^{0}$, Eq.~(\ref{JO}).
It is thus urgent to have a precision measurement on $\Omega^{0}_{3}$, although the task can be
very challenging.

Turning to neutrino propagation in matter, we find that the parameters have simple
dependences on $A$, the induced mass.  To a good approximation, there are
two resonance regions ($A \sim A_{l}$ and $A \sim A_{h}$) where they change rapidly.
Outside of these there are three regions of $A$ in which all parameters take on 
values that are nearly constant.  These are given in detail in Sec. V.  Starting from
$(X^{0}_{i},\Omega^{0}_{i})$ for vacuum ($A=0$), the list of free parameters gets shorter as $A$ increases.
For $A_{l} \lesssim A \lesssim A_{h}$, there are only two:
$\Omega_{2}(A_{i}) \cong \Omega^{0}_{2}-\Omega^{0}_{1}$, $X_{3}(A_{i}) \cong X^{0}_{3}$.
For $A \gtrsim A_{h}$, it is down to one: $\Omega_{3}(A_{d}) \cong -(\Omega^{0}_{2}-\Omega^{0}_{1})$.
Physically, this behavior is a consequence of decoupling, and can be tested in LBL experiments.
So far, the analysis is done under the assumption of normal hierarchy.
For the case of inverted hierarchy, similar results are obtained with the
omission of the higher resonance.

In conclusion, given the incomplete knowledge that is now available,
it is seen that, with the help of consistency conditions and the approximations 
$X^{0}_{3} \ll 1$ and $\Delta^{0}_{21} \ll \Delta^{0}_{31}$, much can already be learned
quantitatively about $W_{\nu}$, both in vacuum and in matter.  It is hoped that
our analysis will be helpful toward establishing a comprehensive specification
of the neutrino mixing matrix.

\acknowledgments                 
SHC is supported by the National 
Science Council of Taiwan, Grant No. NSC 100-2112-M-182-002-MY3.

\end{document}